\documentstyle[12pt]{article}
\begin{document}

\centerline{\large\bf Influence of Pressure on Smectic A-Nematic Phase
Transition} 
\vskip 0.5in

\centerline{Prabir K. Mukherjee$^1$ and Kisor Mukhopadhyay$^2$}

\vskip 0.2in
\centerline{$^1$Saha Institute of Nuclear Physics, LTP Division,}
\centerline{1/AF Bidhan Nagar, Calcutta 700064, INDIA}
\vskip .1in
\centerline{$^2$Department of Physics, Jadavpur University,}
\centerline{ Calcutta 700032, INDIA} 
\vskip 0.5in

\begin{abstract}
We propose a Landau-de Gennes phenomenological model to describe the
pressure induced smectic A-nematic phase transition. The pressure induced
smectic A phase transitions are discussed for varying coupling between
orientational and translational order parameter. Increasing the
pressure, the first order nematic-smectic A transition becomes second
order at a tricritical point which agrees fairly with available
experimental results.
\end{abstract}
\vskip .5in
{\bf PACS Number(s) :} 42.70. Df, 61.30. 
\newpage
\section{Introduction}
\vskip .1in
In spite of considerable research activity over the past twenty years the
nematic to smectic A (NA) phase transition remains one of the principal
unsolved problems in equilibrium statistical physics. During the past
twenty years, many high resolution heat capacity and x-ray studies have
been devoted to the NA transition [1-11]. The theory of the NA
transition has also received considerable attention [12-20]. At the NA
phase transition the continuous translational symmetry of the nematic
phase is sponteneously broken by the appearence of  one dimensional
density wave in the smectic A phase. Original theories due to McMillan
[13] and de Gennes [14] suggested that NA transition could be first or
second order. The order of the transition changes at a tricritical point
(TCP). Alben [21] predicted a $^3$He-$^4$He-like TCP in binary
liquid-crystal mixtures. However, Halperin, Lubensky and Ma [15] argued
that NA transition can never be truly second order, which ofcourse rules
out the possibility of a TCP. This controversy has spurred experimental
studies [1-11] which have shown that NA transition can indeed be
continuous when measured to the dimensionless temperature
$(T-T_{NA})/T_{NA}$ $\approx$ 10$^{-5}$. Also, the Landau and molecular
field theories for the tricritical point are discusses by Longa [22]. In a
very recent paper [19] Lelidis and Durand predicted the field induced TCP
in NA transition.

Again a pressure induced TCP of NA transition was studied
experimentally by Mckee et al. [23]. They examined the effects of pressure
on the NA transition by measuring a quantity proportional to the nematic
order parameter, a splitting $\Delta H$ in the proton dipolar NMR spectrum,
at pressure up to 3.5 kbar. They pointed out that the confirmation of a NA
TCP at elevated pressure would suggest the existence of compounds with a
second order transition. Again, Keyes, Weston and Daniels [24] measured
turbidity on Cholesteryl oleyl carbonate, reported evidence of a change in
the nature of the cholesteric-smectic A phase transition under pressure
suggestive of a TCP.

The purpose of the present paper is to examine the effects of pressure on NA
transion within Landau-de Gennes phenomenological theory. In this analysis
we find a pressure induced TCP on NA transition as suggested by [23].
\vskip .1in

\section{Model}
\vskip .1in
We start by defining an order parameter for the smectic A phase. The order
parameter, $\psi$, is defined as the amplitude of a one dimensional
density wave whose wave vector, $q_0$, is parallel to the director (the
Z-axis) :
\begin{equation}
\rho(\vec r) = \rho_0 \{ 1 + {\rm Re} [\psi e^{i q_0 z}] \} ,
\end{equation} 
where $a = 2 \pi /q_0$ is the layer spacing and $\psi(\vec r) = |\psi|
e^{i q_0 u}$. Here $u$ is the displacement of the layers in the
Z-direction away from their equilibrium position (k=k(x,y,z)). We imagine
that there exists a pressure which decreases the free energy density F.
Hence the free energy density of the NA transition in the presence of
pressure written as
\begin{equation}
F = F_N(T,Q) + F_A (T,\psi) + F_{AN} (Q,\psi) + F_P(Q,\psi,P) ,
\end{equation}
where $F_N(T,Q)$ is the free energy density of the nematic phase,
$F_A(T,\psi)$ corresponds to the smectic A phase, $F_{AN}$ is the
contribution from the coupling between $Q$ and $\psi$ and $F_P(Q,\psi,P)$
is the free energy density associated with the coupling of the pressure
with the order parameters. Now $F$ can be expanded in powers of nematic
order parameter $Q$ and smA order parameter $\psi$ :
$$
F_N = F_0 + \frac{1}{2} A Q^2 + \frac{1}{3} B Q^3 + \frac{1}{4} C Q^4 ,
\eqno{(3a)}
$$
where, $A = a(T-T_{NI}^{\ast})$, $F_0$ is the free energy density of the
isotropic phase and $T_{NI}^{\ast}$ is the isotropic supercooling
temperature. $a$, $B$ and $C$ are independent of temperature. Because of
the first order nematic-isotropic (NI) transition, $a$>0, $B$<0, and $C$>0.
$$
F_A = \frac{1}{2} \alpha |\psi|^2 + \frac{1}{4} \beta |\psi|^4 , \eqno{(3b)}
$$
where $\alpha = \alpha_0(T - T_{AN}^{\ast})$, $T_{AN}^{\ast}$ is the
supercooling temperature limit of the nematic phase for $F_{AN}$=0 and
$P$=0. $\alpha_0$ and $\beta$ are constants. In the absence of coupling
equation (3b) describes a second order transition. Since smectic phase
cannot occur without orientational ordering, hence to ensure the smA phase
$F_{AN}$ is included in expression (2). At lowest order coupling term
$F_{AN}(Q,\psi)$ can be written as follows :
$$
F_{AN} = \gamma Q |\psi|^2 + \frac{\eta}{2} Q^2 |\psi|^2  , \eqno{(3c)}
$$
$\gamma$ and $\eta$ are coupling constants. $\gamma$ is chosen negetive to
favour smA phase when the nematic phase exists and we choose $\eta$>0.
Generally this term allows reentrant effects [25-27]. Because of
($Q$,$\psi$) coupling the NA phase transition can be of second order or
first order [14].

We discuss now the effect of pressure on NA transition. This leads to
describe the coupling of the order parameters with external pressure. From
symmetry consideration, the lowest order term $F_P(Q,\psi,P)$ can be
written as

$$
F_P = \lambda P^2 Q . \eqno{(3d)}
$$

Minimization of (2)corresponds to the equations :
\setcounter{equation}{3}
$$
\begin{array}{rl}
F^{\prime}_{Q} = \frac{\partial F}{\partial Q} &= A Q + B Q^2 + C Q^3 
+ \gamma |\psi|^2 +\lambda P^2 + \eta |\psi|^2 Q = 0 , \\
\end{array}
\eqno{(4a)}
$$
$$
\begin{array}{rl}

\frac{\partial F}{\partial {|\psi|}} &= (\alpha + \beta |\psi|^2 + 2 \gamma
Q + \eta Q^2) |\psi| = 0 .\\
\end{array}
\eqno{(4b)}
$$
The stability conditions are 
$$
\begin{array}{rl}

F^{\prime \prime}_{Q} = \frac{\partial^2 F}{\partial Q^2} &= A + 2 B Q + 3
C Q^2 + \eta |\psi|^2 > 
0 ,\\
\end{array}
\eqno{(5a)}
$$
$$
\begin{array}{rl}

\frac{\partial^2 F}{\partial |\psi^2|} &= \alpha + 3 \beta |\psi|^2 + 2
\gamma Q + \eta Q^2 > 0 .\\
\end{array}
\eqno{(5b)}
$$
Now in addition to the high symmetry phase (isotropic phase) denoted as I
one finds two other possible stable phases corresponding to (in the
absence of pressure)\\
II) Q $\neq$ 0, \hskip .1in $|\psi|$ = 0 \hskip .3in nematic phase , \\
III) Q $\neq$ 0, \hskip .1in $|\psi|$ $\neq$ 0 \hskip .3in
smectic phase.\\
Now from equation (4b) we find
$$
|\psi|^2 = \left \{
\begin{array}{ll}
\frac{-(\alpha + 2 \gamma Q + \eta Q^2)}{\beta} & {\rm if\ } \alpha+2
\gamma Q+\eta Q^2 <0 \\
0 & {\rm if\ } \alpha+2\gamma Q+\eta Q^2 > 0
\end{array}
\right.
\eqno{(6)}
$$ 
Then the equation 
\setcounter{equation}{6}
\begin{equation}
\alpha + 2 \gamma Q + \eta Q^2 = 0
\end{equation}
indicate the existence of sm A phase. It is also clear from equation (7)
that the transition to the sm A phase depends on the temperature
$T_{AN}^{\ast}$ and on the ratio $\gamma/\alpha_0$.

Let us now recall the properties of NI transition. Since the nematic phase
has only orientational order, equation (4a) simplifies to
$$
F_{Q}^{\prime} = A Q + B Q^2 + C Q^3 + \lambda P^2 = 0 \eqno{(8a)}
$$
and the stability condition (5a) gives :
$$
F_{Q}^{\prime \prime} = A + 2 B Q + 3 C Q^2 > 0 . \eqno{(8b)}
$$
Inside this stability region the system is completely unstable. Equation
(8a) describes the first order NI transition at the temperature $T_{NI}$
in the absence of pressure. In the presence of pressure the NI transition
becomes second order at the isolated critical point in the
temperature-pressure diagram. When pressure acts on the NI transition, the
minimum corresponding to isotropic state is no more $Q$=0. It is shifted
to a small but non-zero value, proportional to $P^2$. Then at the isolated
critical point NI transition becomes second order.
The existence of this isolated critical point was verified by several
authors [28-31]. The coordinates of the isolated critical point are
\setcounter{equation}{8}
\begin{eqnarray}
\nonumber
Q_c &=& -\frac{B}{3 C},\\
P_{c}^{2} &=& \frac{B^3}{27 \lambda C^2},\\
\nonumber
T_c &=& T_{NI}^{\ast} + \frac{\beta^2}{3 a C} .
\end{eqnarray}
Now to find out the variation of the pressure in the sm A phase let us now
recall the solution $Q\neq$0, $|\psi| \neq $ 0. For $|\psi| \neq$ 0, $F$
can be written as function of $Q$ alone. By substituting $|\psi (Q)|^2$
from equation (6) into equation (2) we obtain :
\begin{equation}
F = F_{0} - \frac{\alpha^2}{4 \beta} + (\lambda P^2 - \frac{\gamma
\alpha}{\beta} ) Q + \frac{1}{2} A^{\ast} Q^2 + \frac{1}{3} B^{\ast} Q^3 +
\frac{1}{4} C^{\ast} Q^4 ,
\end{equation}
where the renormalized coefficients are :
$$
\begin{array}{rl}
A^{\ast} &= A - (2 \gamma^2 + \eta \alpha)/\beta ,\\
\end{array}
\eqno{(11a)}
$$
$$
\begin{array}{rl}
B^{\ast} &= B - 3 \eta \gamma /\beta ,\\
\end{array}
\eqno{(11b)}
$$
$$
\begin{array}{rl}
C^{\ast} &= C - \eta^2/\beta .\\
\end{array}
\eqno{(11c)}
$$
The equilibrium condition of equation (10) can be written as 
\setcounter{equation}{11}
\begin{equation}
(\lambda P^2 - \frac{\gamma \alpha}{\beta}) + A^{\ast} Q + B^{\ast} Q^2 +
C^{\ast} Q^3 = 0 .
\end{equation}
Here we assume that  $C > \eta^2/\beta$. When $C$ < $\eta^2 \beta$ then a
higher order term should be included. Let us now assume $Q_{NA}$ be the
nematic order parameter at the NA transition and $F_N(Q_{NA})$ is the free
energy of the nematic phase at the NA phase transition point and
$F^{\prime \prime}_{Q_{NA}}$ is the second derivative of $F_N(Q_{NA})$ at
NA point. Then the free energy $F$ of the smectic phase can be written as,
\begin{equation}
F \approx F_N(Q_{NA}) + \frac{1}{2} F_{Q_{NA}}^{\prime \prime}(Q -
Q_{NA})^2 + \lambda P^2 Q + \frac{1}{2} \alpha |\psi|^2 + \frac{1}{4}
\beta |\psi|^4 + \gamma Q |\psi|^2 .
\end{equation}
Minimization of (13) gives
\begin{equation}
F_{Q_{NA}}^{\prime \prime} (Q - Q_{NA}) + \lambda P^2 + \gamma |\psi|^2 = 0.
\end{equation}
Hence from equation (14) 
\begin{equation}
Q = -\frac{(\lambda P^2 + \gamma |\psi|^2)}{F^{\ast}_{Q_{NA}}} +
\frac{Q_{NA}}{F^{\prime \prime}_{Q_{NA}}} .
\end{equation}
Substitution of equation (15) in equation (13) gives 
\begin{equation}
F \approx F_{N}(Q_{NA},P) + \frac{1}{2} \alpha^{\ast} |\psi|^2 +
\frac{1}{4} \beta^{\ast} |\psi|^4 .
\end{equation}
Here,
$$
\begin{array}{rl}
\alpha^{\ast} &= \alpha - \frac{2 \gamma \lambda P^2}{F^{\prime
\prime}_{Q_{NA}}} + 2 \gamma Q_{NA} ,\\
\end{array}
\eqno{(17a)}
$$
$$
\begin{array}{rl}
\beta^{\ast} &= \beta - \frac{2 \gamma^2}{F^{\prime \prime}_{Q_{NA}}} .\\
\end{array}
\eqno{(17b)}
$$
Hence the renormalization of the transition temperature due to the ($Q$,
$\psi^2$) coupling and the pressure effect can be written as
\setcounter{equation}{17}
\begin{equation}
\tilde T_{NA}^{\ast} = T_{NA}^{\ast} + \frac{2 \gamma \lambda
P^2}{F^{\prime \prime}_{Q_{NA}}} + 2 \gamma Q_{NA} .
\end{equation}
From equation (17b) we see that for the strong enough coupling
$\beta^{\ast}$ can become negative i.e. the NA transition becomes first
order. To ensure stability a positive stabilizing sixth order term should
be added in the free energy expansion (16). For the weak coupling,
$\beta^{\ast}$ remains positive and the transition becomes second order.
When ($Q$, $\psi$) coupling is vanishingly weak the only influence of the
pressure on the smectic order comes from the coupling between smectic
order parameters and pressure. Thus the free energy (13) approximated as
\begin{equation}
F \approx F_N(Q_{NA}) + \frac{1}{2} F^{\prime \prime}_{Q_{NA}} (Q -
Q_{NA})^2 + \lambda P^2 Q + \frac{1}{2} \alpha |\psi|^2 + \frac{1}{4}
\beta |\psi|^4 + \frac{1}{2} \delta P^2 |\psi|^2 .
\end{equation}
After minimization and eliminating $Q$, $T_{AN}^{\ast}$ is renormalized to
\begin{equation}
T_{AN} (\gamma = 0) = T_{AN}^{\ast} - \frac{\delta}{\alpha_0}P^2 .
\end{equation}
In the absence of coupling, as $\delta$/$\alpha_0$ < 0, the ($\psi$,$P$)
coupling increases the transition temperature proportional to the square of
the pressure, but it does not influence the order of the transition and
the two transitions in $Q$ and $\psi$ have been chosen stable. Now one
calculates the pressure induced NA transition temperature $T_{AN}(P)$ and
the order $Q_{AN}(P)$ at the transition point numerically from  equations
(7) and (8a). To do this calculation we have to calculate the value of
coefficients. Since these value are still unknown, we are not in a
position to calculate $T_{AN}(P)$ and $Q_{AN}(P)$ numerically. One
concludes that $T_{AN}(P)$ is increasing with the pressure as expected. As
the nematic order $Q_{NA}$ at the transition increases with the field :
$Q_{NA} \rightarrow Q_{NA} - \frac{\lambda P^2}{F^{\prime
\prime}_{Q_{NA}}}$ the appearence of the smectic order at the transition
has smaller influence on the nematic order.

In the absence of pressure, the NA transition at $t_{AN}$ (= $T_{NA} -
T_{NI}^\ast$)  is of first order and shows a jump of $Q$ because because
positional order appears discontinuously. The transition temperature
$T_{AN}(P)$ is defined from the conditions
$$
F (Q \neq 0, |\psi| \neq 0) = (Q \neq 0, |\psi| = 0).
$$
The intersection of the smectic spinodal with the equilibrium line $Q(T,
P=0)$ gives the superheating limit $T_{AN}^{\ast \ast}$ of the first order
NA transition in absence of pressure. The intersection of the spinodal
with the smectic line defines the TCP where the transition becomes second
order with coordinates ($Q_{tcp}$,$T_{tcp}$) given by,
$$
\begin{array}{rl}
Q_{tcp} &= - \frac{(B^\ast - a^\ast \gamma/\alpha_0)}{(3 C^\ast - a^\ast
\eta/\alpha_0)} + \sqrt{ \left ( \frac{B^{\ast} - a^\ast \gamma /\alpha_0}{3
C^\ast - a^\ast \eta/\alpha_0} \right )^2 - \frac{a^\ast (T_{AN}^{\ast} -
T^{\ast})}{3 C^{\ast} - a^{\ast} \eta/\alpha_0}} ,\\
\end{array}
\eqno{(21a)}
$$
$$
\begin{array}{rl}
T_{tcp} &= T_{AN}^{\ast} - Q_{tcp}(2 \gamma + \eta Q_{tcp})/\alpha_0 .\\
\end{array}
\eqno{(21b)}
$$
where 
$$
\begin{array}{rl}
a^{\ast} = a_0 - \frac{\eta \alpha_0}{\beta} ,
\end{array}
\eqno{(21c)}
$$
and 
$$
\begin{array}{rl}
T^\ast =
(\frac{a_0 T_{NI}^{\ast}}{a^\ast}) + (2 \gamma^2 - \frac{\eta \alpha_0
T_{AN}^{\ast}}{\beta a^\ast}) .
\end{array} 
\eqno{(21d)}
$$
Thus the tricritical point is
($T_{tcp}$,$Q_{tcp}$). The "tricritical pressure" to attain the TCP is
given by
\setcounter{equation}{21}
\begin{equation}
P_{tcp}^{2} = \frac{1}{\lambda} \left [\frac{\alpha T_{tcp} \gamma}{\beta}
- \alpha^{\ast} T_{tcp} Q_{tcp} - B^{\ast} Q^{2}_{tcp} - C^{\ast}
Q^{3}_{tcp} \right ] .
\end{equation}
Thus for $P$ = $P_{tcp}$ the TCP appears and the transition becomes second
order. Thus we point out that the confirmation of a nematic-smectic A TCP
at elevated pressure would suggest the existence of second order
transition which support the experimental evidences [23]. 

Again from equation (21d) we find that 
\begin{equation}
\frac{dT_{AN}^{\ast}}{dP} = \frac{a_0 \beta}{\eta \alpha_0} 
\frac{dT_{NI}^{\ast}}{dP} ,  
\end{equation}
we get
\begin{equation}
\frac{dT_{NI}^{\ast}}{dP} < \frac{dT_{AN}^{\ast}}{dP} .
\end{equation}
Equation (24) agrees with the experimental observation [23].
\section{Conclusion}
We have developed a phenomenological model in the Landau-de Gennes
formalism to describe the pressure induced phase transitions between
isotropic, nematic and smectic A. The effect of pressure on the second
order NA transition is to increase the transition temperature as square of
the pressure. In presence of pressure, the first order NA transition (when
pressure reduced to zero) becomes of second order above at TCP. The
coordinates TCP have been calculated. The value of the pressure at TCP
have also been calculated. Thus the present model analyze the experimental
results [23].
\vskip .3in
\noindent{\bf Acknowledgement}\\
We are grateful to Dr. Soumen Kumar Roy for his valuable comments 
on an earlier version of this paper. One of the author (KM) wishes to thank the 
{\it Council of Scientific and Industrial Research, Government of India} for 
the award of senior research fellowship.
\end{document}